\documentclass[aps,prb,reprint,floatfix,preprintnumbers,showpacs,citeautoscript,
superscriptaddress]{revtex4-1}
\usepackage{graphicx}
\usepackage{amsfonts,amssymb,amsmath}
\usepackage[utf8x]{inputenc}
\usepackage{xcolor}
\usepackage{etoolbox}

\apptocmd{\sloppy}{\hbadness 10000\relax}{}{}

\newcommand{\bo}{\mathbf}
\newcommand{\bs}{\boldsymbol}
\newcommand{\da}{\downarrow}
\newcommand{\eff}{\text{eff}}
\newcommand{\imp}{\text{imp}}
\newcommand{\pdag}{\phantom{\dag}}

\newcommand{\ua}{\uparrow}

\newcommand{\Aimp}{A_{\imp}}
\newcommand{\chiimp}{\chi_{\imp}}
\newcommand{\Simp}{S_{\imp}}

\begin{document}

\title{Kondo effect in graphene with Rashba spin-orbit coupling}
\date{\today}

\author{D.\ Mastrogiuseppe}
\affiliation{Department of Physics and Astronomy, and Nanoscale and Quantum
Phenomena Institute, \\ Ohio University, Athens, Ohio 45701--2979, USA}
\affiliation{Dahlem Center for Complex Quantum Systems and Fachbereich Physik,
Freie Universit\"at Berlin, 14195 Berlin, Germany}
\author{A.\ Wong}
\affiliation{Department of Physics, University of Florida, P.O.\ Box 118440,
Gainesville, Florida, 32611--8440, USA}
\author{K.\ Ingersent}
\affiliation{Department of Physics, University of Florida, P.O.\ Box 118440,
Gainesville, Florida, 32611--8440, USA}
\author{S.\ E.\ Ulloa}
\affiliation{Department of Physics and Astronomy, and Nanoscale and Quantum
Phenomena Institute, \\ Ohio University, Athens, Ohio 45701--2979, USA}
\affiliation{Dahlem Center for Complex Quantum Systems and Fachbereich Physik,
Freie Universit\"at Berlin, 14195 Berlin, Germany}
\author{N.\ Sandler}
\affiliation{Department of Physics and Astronomy, and Nanoscale and Quantum
Phenomena Institute, \\ Ohio University, Athens, Ohio 45701--2979, USA}
\affiliation{Dahlem Center for Complex Quantum Systems and Fachbereich Physik,
Freie Universit\"at Berlin, 14195 Berlin, Germany}

\begin{abstract}
We study the Kondo screening of a magnetic impurity adsorbed in graphene in
the presence of Rashba spin-orbit interaction. The system is described by an
effective single-channel Anderson impurity model, which we analyze using the
numerical renormalization group. The nontrivial energy dependence of the host
density of states gives rise to interesting behaviors under variation of the
chemical potential or the spin-orbit coupling. Varying the Rashba coupling
produces strong changes in the Kondo temperature characterizing the many-body
screening of the impurity spin, and at half-filling allows approach to a
quantum phase transition separating the strong-coupling Kondo phase from a
free-moment phase. Tuning the chemical potential close to sharp features of
the hybridization function results in striking features in the temperature
dependences of thermodynamic quantities and in the frequency dependence of
the impurity spectral function.
\end{abstract}


\maketitle

\section{Introduction}
\label{sec:intro}

Two-dimensional ``single-layer'' materials have generated great attention over
the past decade \cite{Novoselov2005,CastroNeto2009}. Understanding the physics
of these low-dimensional systems is highly desirable for the design of a new
generation of electronic devices with on-demand characteristics. Single- or
few-layer structures facilitate control by chemical and/or electrical means, as
well as direct access to electronic features using local probes such as scanning
tunneling microscopy \cite{WiesendangerBook}.

Coupling between orbital and spin degrees of freedom has noticeable effects on
the properties of many single-layer materials. Spin-orbit interaction (SOI) can
come into play either in its intrinsic form arising from the presence of
constituent atoms with high atomic number, for instance in transition-metal
dichalcogenides, or in extrinsic form due to the breaking of the electron spin
degeneracy through spatial-inversion asymmetry arising from a substrate or an
external electric field. The control and manipulation of SOI is also important
for spintronics applications \cite{Yazyev2010}.

One fascinating direction relates to the physics of localized magnetic moments
on single-layer structures and their collective screening through the Kondo
effect \cite{Hewson}. This topic is of particular interest in the context of
graphene, the prototypical single-layer material and perhaps the most basic
condensed-matter system to feature low-energy Dirac fermions. Many analytical
and a few experimental studies have recently reported controversial and
sometimes conflicting results \cite{Fritz2013,Jobst2013} concerning the
existence and nature of the Kondo effect. Among other suggestions has been
the exciting possibility of accessing a regime of non-Fermi liquid
multichannel Kondo physics \cite{Manoharan, Sengupta2008, Kharitonov2013}. 

The goal of this paper is to advance understanding of the Kondo physics of a
magnetic impurity adsorbed on graphene, properly taking into account Rashba SOI.
Each of the important ingredients of this problem has been considered
previously, but hitherto they have not all been treated within a single
setting.

The effect of Rashba SOI on Kondo screening in conventional two-dimensional
electron gases has been analyzed via a number of methods in the recent
literature \cite{Malecki2007,Zitko2011,Feng2011,Zarea2012,Isaev2012}. It has
been shown, in particular, that the Kondo temperature can be exponentially
enhanced when the chemical potential is tuned to lie close to a van Hove
singularity that arises in the density of states (DOS) due to SOI
\cite{Wongprep}. This is an illustration of how nontrivial structure in the
single-particle DOS can greatly impact the Kondo many-body effect.

In the absence of SOI, it is expected that magnetic-impurity physics in
pristine graphene will reflect a linear vanishing of the DOS around the energy
of the Dirac points, corresponding to the case $r=1$ of a pseudogap DOS in
which $\rho(E)\propto |E-E_F|^r$ for energies $E$ close to the Fermi energy
$E_F$. The pseudogap Kondo and Anderson models (first studied in the context of
unconventional superconductivity, narrow-gap semiconductors, and flux phases)
have interesting phase diagrams that depend on the strength of coupling between
the impurity and the conduction band, as well as the presence or absence of
particle-hole ($p$-$h$) symmetry \cite{Withoff1990, Ingersent1996, BPH97,
GBuxton1998}.

Even though SOI is weak for graphene deposited on conventional substrates, it
has been recently shown \cite{Marchenko2012} that Au intercalation at the
interface between graphene and an Ni substrate can give rise to a spin-orbit
splitting as large as $100$ meV. A large SOI has also been achieved in
hydrogenated graphene at low concentrations, with Rashba coupling in the meV
range \cite{Balakrishnan2013}. The effect of Rashba SOI is to transform the
linear band dispersion around each Dirac point into a four-band hyperbolic
dispersion, generating discontinuities in the DOS at energies depending on the
Rashba coupling parameter. SOI also converts to a nonzero DOS what in the
absence of SOI would be a vanishing value at the charge-neutrality point, while
still preserving a strong energy dependence.

Isolated magnetic moments can be generated in graphene either by decorating the
sample with adatoms or molecules, or with vacancies produced by irradiation
\cite{Ugeda2010}. Most density functional (DFT) studies suggest that transition
metals adatoms tend to prefer adsorption at hollow sites in the center of
hexagons \cite{Mao2008, Sevincli2008,Zanella2008,Johll2009,Krasheninnikov2009,
Cao2010,Valencia2010,Sargolzaei2011,Ding2011,Zhang2013}. However, it has been
shown that the adsorption site is strongly influenced by the value of the local
Coulomb interaction as reported by GGA+U calculations \cite{Wehling2011}.
Additionally, recent experimental works show that the adsorption site for Co
and Ni atoms depends strongly on the substrate: Co always adsorbs onto SiC on
top of a carbon atom, while for freestanding graphene one finds top and hollow
sites for both Co and Ni \cite{EelboA2013,*EelboB2013}. While adsorbed Ni
always seems to be nonmagnetic, Co carries a magnetic moment that depends on
the adsorption site.

In this paper we study an Anderson impurity model describing a configuration in
which a spin-$\frac12$ adatom with axial orbital symmetry is adsorbed on top of
a graphene carbon atom, the most probable adsorption site for Co
\cite{EelboA2013,*EelboB2013}. The hybridization between the magnetic impurity
and graphene bulk states inherits nontrivial energy dependence from the DOS.
Using the numerical renormalization-group (NRG) technique, we calculate
thermodynamic and spectral quantities that enable us to characterize the Kondo
physics under variation of the chemical potential and the Rashba coupling. If
the system is held at half-filling, increasing the Rashba parameter from zero
gives rise to a quantum phase transition. When the system is doped
such that the Fermi level lies close to a discontinuity in the impurity-band
hybridization function, the proximity of a new screening channel produces a
suppression in the Kondo peak near the Fermi energy, as can be seen in the
impurity spectral function. The Kondo temperature, which exhibits a strong
dependence on band filling, undergoes a sudden change as the chemical potential
crosses a jump in the hybridization function. Additionally, the impurity is
found to make a negative low-temperature contribution to thermodynamic
quantities such as the magnetic susceptibility and the entropy. This rich
behavior is in principle accessible in experiments.

The remainder of the paper is organized as follows. Section \ref{sec:model}
describes the model and its transformation into a form that can be solved
using the NRG method. Results are presented in Sec.\ \ref{sec:results},
followed by discussion and our conclusions in Sec.\ \ref{sec:discuss}.

\section{Model and identification of relevant screening channel}
\label{sec:model}

Our analysis is based on an Anderson Hamiltonian for a system in which massless
Dirac fermions of the host graphene experience Rashba SOI and are coupled to a
nondegenerate impurity level exhibiting axial symmetry about a direction
perpendicular to the plane of the graphene. We show in this section how the
resulting multichannel model can be reduced to a single-channel Hamiltonian via
a sequence of canonical transformations on the fermionic degrees of freedom.
These transformations allow the energy-dependent impurity hybridization function
to be identified.

\subsection{Model Hamiltonian}
\label{subsec:model}

We start with a real-space tight-binding Hamiltonian for graphene:
\begin{equation}
\label{H_g}
H_g = -t \sum_{\bo{R}, j, s} a_s^{\dag}(\bo{R}) \:
  b_s^{\pdag}(\bo{R} + \bs{\delta}_j) + \text{H.c.},
\end{equation}
where $a_s(\bo{R})$ [$b_s(\bo{R})$] destroys an electron with spin $z$
projection $\pm\frac12$ for $s=\pm 1$ (alternatively $s=\,\ua\!/\!\da$) on
the sublattice-$A$ [sublattice-$B$] carbon atom in the unit cell centered at
$\bo{R}$. Also, $t\simeq 3$\,eV is the nearest-neighbor hopping, and
$\bs{\delta}_j$ ($j = 1$, $2$, $3$) are nearest-neighbor translation
vectors of length $a\simeq 1.42$\,\AA\ lying in the $x$-$y$ plane.
After an expansion around the two nonequivalent Dirac points, located in
two-dimensional reciprocal space at wave vectors
$\bo{K}_{\pm}=(\pm 4\pi/(3\sqrt{3}a),\,0)$,
we arrive at the low-energy effective Hamiltonian
\begin{equation}
\label{H_g^eff}
H_g^{\eff}
  = \sum_{\bo{q}}  \psi^\dag(\bo{q}) \: h_g(\bo{q}) \: \psi(\bo{q}),
\end{equation}
with (setting $\hbar = 1$)
\begin{equation}
\label{valley_isotrop}
h_g(\bo q) = \tau_0 \otimes s_0 \otimes
  ( v_F \, \bo{q} \cdot \bs{\sigma} - \mu \sigma_0 ),
\end{equation}
where $\bo{q} = \bo{k} - \bo{K}_{\tau}$ is the wave vector measured from the
center of valley $\tau=\pm$, $v_F=3at/2$ is the Fermi velocity, $\mu$ is the
chemical potential, $\bs{\sigma} = (\sigma_x,\sigma_y)$ is a vector of Pauli
matrices acting on a sublattice pseudospin degree of freedom $\sigma=\pm 1$,
and $\tau_0$, $s_0$, and $\sigma_0$ are the two-dimensional identity matrices
in the valley, spin, and pseudospin spaces, respectively. The Hamiltonian
matrix \eqref{valley_isotrop}, written in a valley-isotropic representation
\cite{Beenakker2008} that allows one to treat both valleys on the same footing,
was obtained by arranging the eight-component spinor $\psi(\bo{q})$ as
\begin{equation}
\label{psi_def}
\psi(\bo{q}) = \bigl( \psi_{+,\ua}(\bo{q})^T, \: \psi_{+,\da}(\bo{q})^T, \:
               \psi_{-,\ua}(\bo{q})^T, \: \psi_{-,\da}(\bo{q})^T \bigr)^T,
\end{equation}
with $\psi_{+,s} = (a_{+,s} \, , \, b_{+,s})^T$ and
$\psi_{-,s} = (b_{-,s} \, , \, -a_{-,s})^T$ being indexed by $\sigma=1$
or $-1$. Here, $a_{\tau,s}(\bo{q})$ [$b_{\tau,s}(\bo{q})$] is the
annihilation operator in sublattice $A$ [$B$], valley $\tau$,
spin $z$ projection $s/2$, and relative wave vector $\bo{q}$.

The Rashba term, arising from asymmetry of the potential along the
$z$ axis, is written in the tight-binding representation as
\cite{Kane2005}
\begin{equation}
\label{H_R}
H_R = i \frac{\lambda_R}{a} \sum_{\bo{R}, j, s, s'} a_s^{\dag} (\bo{R}) \,
  [\bo{s}_{ss'}\times\bs{\delta}_j]_z \: b_{s'}^{\pdag}(\bo{R}+ \bs{\delta}_j)
  + \text{H.c.},
\end{equation}
where $s_x$ and $s_y$ are Pauli matrices in the spin space.
A low-energy reciprocal-space representation can be deduced through expansion
of Eq.\ \eqref{H_R} about each Dirac point, or it can be introduced by
symmetry arguments \cite{Kane2005}. To zeroth order in $q_x a$ and $q_y a$,
\begin{equation}
\label{H_R^eff}
H_R^{\eff} = \sum_{\bo{q}}  \psi^{\dag}(\bo{q}) \, h_R \, \psi(\bo{q}),
\end{equation}
with
\begin{equation}
\label{h_R}
h_R = \lambda \, \tau_0 (\bs{s} \times \bs{\sigma}) \cdot \hat{\bo{z}}
    = \lambda \, \tau_0 \left(s_x \sigma_y - s_y \sigma_x\right) .
\end{equation}
where the sign and magnitude of the Rashba parameter $\lambda=3\lambda_R/2$
can be tuned in experiments via an electric field applied
parallel or antiparallel to the $z$ axis.
For convenience, we take $\lambda\ge 0$. 

We consider a magnetic impurity level that adsorbs to the sublattice-$A$
carbon atom in the unit cell at $\bo{R}=\bo{0}$. The nondegenerate level
of the isolated impurity atom can be described by
\begin{equation}
\label{H_imp}
H_{\imp} = \epsilon_d \sum_s n_{ds} + U n_{d\uparrow} n_{d\downarrow},
\end{equation}
where $\epsilon_d$ is the impurity level energy relative to the chemical
potential, $U$ is the onsite Coulomb repulsion, and $n_{ds}= d_s^{\dag}
d_s^{\pdag}$ with $d_s$ destroying an electron of spin $s$ in the impurity 
level. Assuming that electron tunneling to/from the impurity level takes
place only through the nearest carbon $p_z$ orbital, mixing between the impurity 
and the host is captured in a hybridization Hamiltonian term 
\begin{equation}
H_{\text{mix}} = V \sum_{s} d_s^{\dag} \: a_s^{\pdag}(\bo{0}) + \text{H.c.},
\end{equation}
which, after the Fourier transformation and expansion in reciprocal space
about the Dirac points yields
\begin{equation}
\label{H_mix}
H_{\text{mix}} = \frac{V}{\sqrt{N_c}} \sum_{\tau,s,\bo{q}}
   d_s^{\dag} \: a_{\tau,s}^{\pdag}(\bo{q}) + \text{H.c.},
\end{equation}
where $N_c$ is the number of unit cells in the graphene layer.

Our goal is to understand the low-energy physics of the Hamiltonian
$H = H_g^{\eff} + H_R^{\eff} + H_{\imp} + H_{\text{mix}}$ defined
in Eqs.\ \eqref{H_g^eff}--\eqref{psi_def}, \eqref{H_R^eff}--\eqref{H_imp},
and \eqref{H_mix}. In order to accomplish this, it is helpful to perform
several simplifications described in the next subsection.

\subsection{Transformation of the model}

Taking advantage of the axial symmetry of the impurity about the $z$ axis
perpendicular to the graphene plane, it is convenient to expand in an
angular momentum basis\cite{Malecki2007}
\begin{align}
a_{\tau,s}(\bo q) = \frac{1}{\sqrt{2\pi q}} \sum_{m=-\infty}^{\infty}
  e^{i m \theta} \: a_{\tau,s}^m(q),
\end{align}
where $q=|\bo{q}|$, $\theta=\arctan (q_y/q_x)$, $m$ is the azimuthal quantum
number, and the prefactor of the summation ensures the anticommutation of
the new operators. With a similar expression for the operator $b_{\tau,s}$,
it is convenient to define new 8-component spinors
\begin{equation}
\psi_j(q) = \bigl( \psi_{+,\ua}^j(q)^T, \: \psi_{+,\da}^j(q)^T, \:
   \psi_{-,\ua}^j(q)^T, \: \psi_{-,\da}^j(q)^T \bigr)^T
\end{equation}
with $\psi_{+,s}^j = \bigl( a_{+,s}^{j-s-1/2},\,b_{+,s}^{j-s+1/2} \bigr)^T$  
and  $\psi_{-,s}^j = \bigl( b_{-,s}^{j-s-1/2},$ $-a_{-,s}^{j-s+1/2} \bigr)^T$.
Since each operator entering $\psi_j$ satisfies $m+\frac12(\tau\sigma+s)=j$,
this spinor acts to decrease by $j$ the total angular momentum defined as
$J_z = L_z + \frac12(\tau_z\sigma_z+s_z)$.

\begin{figure}[bt]
\includegraphics[width=0.47\textwidth]{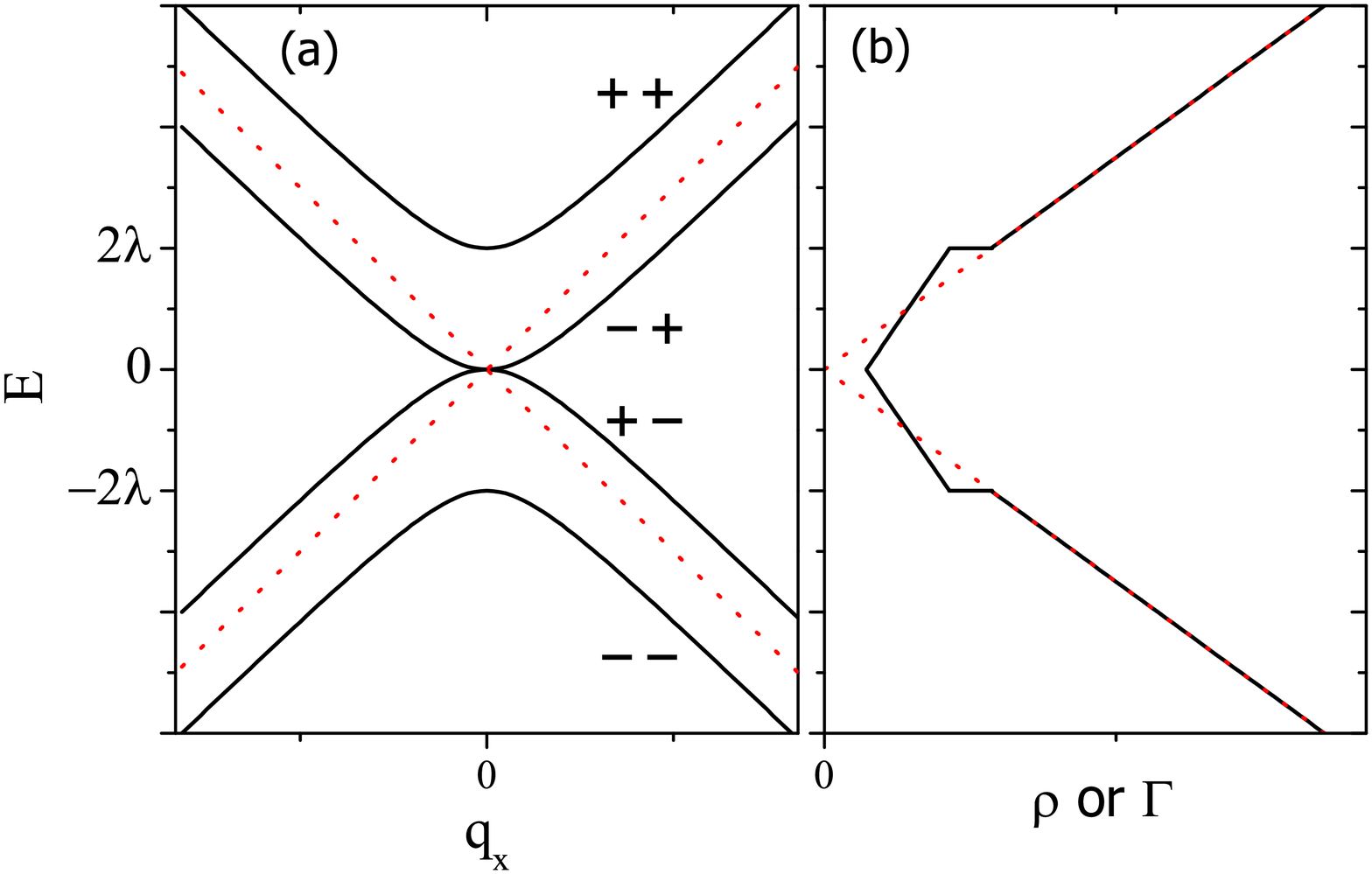}
\caption{\label{fig:dos} (Color online)
(a) Graphene band structure near either Dirac point, plotted schematically as
a function of $q_x$ for fixed $q_y=0$, where $\bo{q}=\bo{k}-\bo{K}_{\pm}$.
Solid [dotted] lines represent the dispersion with [without] Rashba SOI.
Each solid line is labeled with its $\alpha$, $\beta$ values.
(b) The corresponding densities of states $\rho(E)$. For an impurity adsorbed
directly on top of a carbon atom, the hybridization function $\Gamma(E)$ is
simply proportional to $\rho(E)$.}
\end{figure}
  
With integration over $\theta$, the angular component of $\bo{q}$, the bulk
Hamiltonian $H_0^{\eff} \equiv H_g^{\eff} + H_R^{\eff}$ can be rewritten
\begin{equation}
H_0^{\eff} = \sum_{j=-\infty}^{\infty} \int \!\! dq \; \psi_j^{\dag}(q) \;
   [ \tau_0 s_0 ( v_F q \, \sigma_x - \mu \sigma_0 ) + h_R ] \;
   \psi_j^{\pdag}(q) ,
\end{equation}
with $h_R$ still given by Eq.\ \eqref{h_R}. This Hamiltonian can be put into
the diagonal form
\begin{equation}
H_0^{\eff} = \sum_{\alpha,\beta,\tau,j}\int\! dq\:\bigl[ E_{\alpha,\beta}(q)
   - \mu \bigr] \, [f^j_{\alpha,\beta,\tau}(q)]^{\dag} \,
   f^j_{\alpha,\beta,\tau}(q) ,
\end{equation}
where $\alpha$ and $\beta$ run independently over $\pm$, and
\begin{equation}
\label{dispersion}
E_{\alpha,\beta}(q) = \alpha \lambda + \beta \sqrt{(v_F q)^2 + \lambda^2}.
\end{equation}
The dispersions $E_{\alpha,\beta}(q)$ are shown schematically in Fig.\
\ref{fig:dos}(a), along with the corresponding dispersions for $\lambda=0$.
Interestingly, the $\lambda\ne 0$ band structure is similar to that in
Bernal-stacked bilayer graphene \cite{Castro2010} with
$\lambda \rightarrow t_\perp/2$ (where $t_{\perp}$ is the interlayer hopping),
although in the present case the presence of SOI also generates a nontrivial
spin structure \cite{Rashba2009}. The DOS for each $(j,\tau)$ sector is
\begin{equation}
\label{dos}
\rho(E) = \frac{\Omega_0}{\pi v_F^2} \bigl[ |E| + \lambda
   + (|E| - \lambda) \: \Theta(|E|-2\lambda) \bigr],
\end{equation}
where $\Theta(x)$ is the Heaviside step function. This DOS has a linear energy
dependence with discontinuities at $E = \pm 2\lambda$, as shown schematically
in Fig.\ \ref{fig:dos}(b).

We note that terms of higher order in $q_x a$ and $q_y a$ than those contained
in Eqs.\ \eqref{H_g^eff} and \eqref{H_R^eff} modify Eqs.\ \eqref{dispersion}
and \eqref{dos}, both at energies $|E|\gtrsim t/10$ too high to play any
essential part in the Kondo physics and below an energy scale proportional to
$\lambda^4/t^3$ (see Ref.\ \onlinecite{Zarea2009}). The effect of the low-energy
departures from Eqs.\ \eqref{dispersion} and \eqref{dos} will be discussed in
Sec.\ \ref{sec:discuss}.

In the angular-momentum basis, the impurity-host hybridization Hamiltonian
becomes
\begin{equation}
\label{hybrid}
H_{\text{mix}} = V\sqrt{\frac{\Omega_0}{2\pi}} \sum_{s} d_s^{\dag} \,
  \int\! dq \: \sqrt{q} \, a_{\tau,s}^0(q) + \text{H.c.},
\end{equation}
where $\Omega_0$ is the graphene unit cell area. It is evident that this
term involves only orbital angular momentum $m=0$, but that when expressed in
terms of total angular momentum eigenstates, the impurity couples to $j = 0$,
$\pm 1$. Moreover, we need to express the operators $a_{\tau,s}^0$ in terms of
the operators $f^j_{\alpha,\beta,\tau}$ that diagonalize $H_0^{\eff}$.

It is straightforward to show that
\begin{equation}
\label{eigenvec}
\begin{split}
a^{m}_{+,\ua}(q) &= -i \sum_{\alpha,\beta}
   \frac{\alpha}{N_{\alpha,\beta}(q)}
      f^{m+1}_{\alpha,\beta,+}(q), \\
b^{m}_{+,\ua}(q) &= -i \sum_{\alpha,\beta}
   \frac{\alpha E_{\alpha,\beta}(q)}{v_F q \, N_{\alpha,\beta}(q)} 
      f^{m}_{\alpha,\beta,+}(q), \\
a^{m}_{+,\da}(q) &= \sum_{\alpha,\beta}
   \frac{E_{\alpha,\beta}(q)}{v_F q \, N_{\alpha,\beta}(q)}
      f^{m}_{\alpha,\beta,+}(q), \\
b^{m}_{+,\da}(q) &= \sum_{\alpha,\beta}
   \frac{1}{N_{\alpha,\beta}(q)}
      f^{m-1}_{\alpha,\beta,+}(q), \\
b^{m}_{-,\ua}(q) &= -i \sum_{\alpha,\beta}
   \frac{\alpha}{N_{\alpha,\beta}(q)}
      f^{m+1}_{\alpha,\beta,-}(q), \\
a^{m}_{-,\ua}(q) &= i \sum_{\alpha,\beta}
   \frac{\alpha E_{\alpha,\beta}(q)}{v_F q \, N_{\alpha,\beta}(q)}
      f^{m}_{\alpha,\beta,-}(q), \\
b^{m}_{-,\da}(q) &= \sum_{\alpha,\beta}
   \frac{E_{\alpha,\beta}(q)}{v_F q \, N_{\alpha,\beta}(q)}
      f^{m}_{\alpha,\beta,-}(q), \\
a^{m}_{-,\da}(q) &= -\sum_{\alpha,\beta}
   \frac{1}{N_{\alpha,\beta}(q)}
      f^{m-1}_{\alpha,\beta,-}(q),
\end{split}
\end{equation}
where
\begin{equation}
N_{\alpha,\beta}(q) =
   \sqrt{2} \left[ 1 +
     \left( \frac{E_{\alpha,\beta}(q)}{v_F q}\right)^2 \right]^{1/2}
\end{equation}
ensures that the $f$ operators obey the canonical anticommutation relations
\begin{multline}
\bigl\{ f^{j}_{\alpha,\beta,\tau}(q),
  f^{j'}_{\alpha',\beta',\tau'}(q')^{\dag} \bigr\} \\
  = \delta_{\alpha,\alpha'} \, \delta_{\beta,\beta'} \,
    \delta_{\tau,\tau'} \, \delta_{j,j'} \, \delta(q-q') .
\end{multline}
Inserting Eqs.\ \eqref{eigenvec} into
Eq.\ \eqref{hybrid}, we get
\begin{equation}
\begin{split}
H_{\text{mix}} =
& \; V\sqrt{\frac{\Omega_0}{2\pi}} \sum_{\alpha,\beta,\tau,s} \tau \,
   (-i\alpha)^{(1+s)/2} \: d_s^{\dag} \int \! dq \:
   \frac{\sqrt{q}}{N_{\alpha, \beta}(q)} \\
& \; \times \left(\frac{E_{\alpha,\beta}(q)}{q}\right)^{(1-\tau s)/2}
   f^{(\tau+s)/2}_{\alpha,\beta,\tau}(q)+ \text{H.c.}
\end{split}
\end{equation}

At this point, it is convenient to change from integration over wave vector
$q$ to integration over energy $E$. We introduce a function
$\Lambda_{\alpha,\beta}(E)
  = \Theta(\beta E - 2\lambda\delta_{\alpha,\beta})$,
which takes the value 1 for any value of $E$ for which there is a value of $q$
such that $E_{\alpha,\beta}(q) = E$, and which is zero otherwise.
Then we can define new annihilation operators
\begin{align}
f^{j}_{\alpha,\beta,\tau}(E)
&= |dq/dE| \: f^{j}_{\alpha,\beta,\tau}(q) \notag \\
&= \sqrt{\frac{\Lambda_{\alpha,\beta}(E) \, |E-\alpha \lambda|}
   {v_F \sqrt{E(E-2\alpha\lambda)}}} \: f^{j}_{\alpha,\beta,\tau}(q) ,
\end{align}
such that
\begin{multline}
\bigl\{ f^{j}_{\alpha,\beta,\tau}(E),
  f^{j'}_{\alpha',\beta',\tau'}(E')^{\dag} \bigr\} \\
  = \delta_{\alpha,\alpha'} \, \delta_{\beta,\beta'} \,
    \delta_{\tau,\tau'} \, \delta_{j,j'} \, \delta(E-E') .
\end{multline}
In the new basis,
\begin{equation}
H_0^{\eff} = \sum_{\alpha,\beta,\tau,j} \int \! dE \: ( E - \mu ) \:
  f^{j}_{\alpha,\beta,\tau}(E)^{\dag} f^{j}_{\alpha,\beta,\tau}(E),
\end{equation}
and
\begin{align}
\label{H_mix:final}
H_{\text{mix}}
&= \frac{V}{2v_F} \sqrt{\frac{\Omega_0}{2\pi}} \! \sum_{\alpha,\beta,\tau,s}
  \!\! \tau \, (-i \alpha)^{(1+s)/2} \, d_s^{\dag} \! \int \!\! dE \:
  \sqrt{\Lambda_{\alpha,\beta}(E)} \notag \\
&\times |E|^{(1-\tau s)/4} \, |E - 2\alpha\lambda|^{(1+\tau s)/4} \,
  f^{(\tau+s)/2}_{\alpha,\beta,\tau}(E) + \text{H.c.} \notag \\
&= \frac{V}{2} \sum_s d_s^{\dag} \int \! dE \: \sqrt{\rho(E)} \:
   g_s(E) + \text{H.c.} ,
\end{align}
where $\rho(E)$ is the density of states defined in Eq.\ \eqref{dos} and
\begin{equation}
\label{gop}
\begin{split}
g_s(E)
&= \sqrt{\frac{\Omega_0}{2\pi v_F^2 \rho(E)}}
   \sum_{\alpha,\beta,\tau} \tau \, (-i\alpha)^{(1+s)/2} \,
   \sqrt{\Lambda_{\alpha,\beta}(E)} \\
&\times |E|^{(1-\tau s)/4} \, |E-2\alpha\lambda|^{(1+\tau s)/4} \,
   f^{(\tau+s)/2}_{\alpha,\beta,\tau}(E)
\end{split}
\end{equation}
satisfying
\begin{equation}
\bigl\{ g_s^{\pdag}(E), g_{s'}^{\dag}(E') \bigr\}
   = \delta_{s,s'} \, \delta(E-E')
\end{equation}
is the annihilation operator for the single effective band or channel of
host electrons that couples to the magnetic impurity.
The Hamiltonian for the host can be rewritten
\begin{equation}
\label{H_0^eff:final}
H_0^{\eff} = \sum_s \int \! dE \: ( E - \mu ) \: g_s^{\dag}(E) \, g_s^{\pdag}(E)
   + \ldots,
\end{equation}
where ``$\ldots$'' describes degrees of freedom that do not couple to the
impurity and which will henceforth be discarded.

Equations \eqref{H_imp}, \eqref{H_mix:final}, and \eqref{H_0^eff:final} 
represent the reduction of the original four-channel Anderson model
defined in Sec.\ \ref{subsec:model} to an effective one-channel Anderson
impurity model having an impurity hybridization function
\begin{equation}
\label{Gamma}
\Gamma(E) = \frac{\pi V^2}{4} \, \rho(E)          
\end{equation}
that is directly proportional to the DOS $\rho(E)$ shown in Fig.\
\ref{fig:dos}(b).\footnote{Similar reductions of multi-channel Anderson
models to effective one-channel models have been performed previously in
connection with Kondo physics in topological insulators\cite{Zitko2011} and
in a two-dimensional electron gas with Rashba coupling
[R.\ Zitko, Phys.\ Rev.\ B \textbf{81}, 241414 (2010)].}
It  should be noted that even though spin is not a good
quantum number in the presence of SOI, the DOS and thehybridization
function entering the effective Anderson model are spin-independent.

The nontrivial energy dependence of the hybridization
function, with linear regions separated by jumps, suggests that the Kondo
physics will exhibit interesting modulations under variation of the
chemical potential $\mu$ and/or the Rashba parameter $\lambda$. As noted in
Sec.\ \ref{sec:intro}, the DOS of graphene without SOI has the pseudogap
form $\rho(E) \propto |E|^r$ with $r=1$. It is well established that for
$\mu=0$ and $r\ge \frac12$, and in the presence of $p$-$h$ symmetry, no Kondo
screening is possible for any value $V$ of the hybridization \cite{GBuxton1998,
Fritz2013}; instead, the system lies in a free-moment phase in which the
ground state contains a free impurity spin entirely decoupled from the host.
As soon as the Rashba SOI is turned on, however, the DOS at $E=0$ acquires
a finite value $\Omega_0\lambda/\pi v_F^2$ [Eq.\ \eqref{dos} and Fig.\
\ref{fig:dos}]. At fixed $\mu = 0$, therefore, we expect the free-moment
phase present for $\lambda=0$ to be replaced for $\lambda>0$ by a
Kondo-screened phase with a Kondo temperature scale that varies exponentially
with $-1/\lambda$. On the other hand, for fixed Rashba coupling, we expect a
rapid change in the Kondo temperature as $\mu$ crosses the discontinuity in
the hybridization function at $E=\pm 2\lambda$. As will be shown in the next
section, these expectations are borne out by numerical calculations that also
reveal other striking behaviors under variation of $\lambda$ and $\mu$.

\section{Numerical Results}
\label{sec:results}

We have performed numerical renormalization-group (NRG) calculations in order to
rigorously test and quantify the qualitative expectations outlined at the end of
Sec.\ \ref{sec:model}. The NRG is a nonperturbative method that allows the
iterative diagonalization of the Hamiltonian for a quantum impurity model and
yields reliable low-energy many-body states \cite{Wilson1975,Bulla2008}.
These states can be used to calculate dynamical properties such as the impurity
spectral function, as well as the impurity contribution to a thermodynamic
property,\footnote{More recent NRG advances\cite{Anders2005,
Weichselbaum2007} allow superior calculation of spectral properties and
thermodynamics in magnetic fields. However, for the present work the standard 
method is satisfactory.}
defined to be $X_{\imp} = X - X_0$ where $X$ is the value of the
property in the full system consisting of the impurity coupled to the host, and
$X_0$ is the corresponding value for the host alone.
 
We adopt units where $\hbar=k_B=g\mu_B=D=1$. All results shown are for the case
of a $p$-$h$-symmetric impurity (i.e., $U=-2\epsilon_d$) with $U=0.02$, and for
a hybridization-function prefactor $\Gamma_0\equiv\Omega_0 D V^2/(4v_F^2)=0.04$
or $0.08$. The data were calculated using an NRG discretization parameter
$\Lambda=2.5$, retaining 2\ 000 many-body states after each iteration.

\subsection{Kondo temperature}

\begin{figure}[tb]
\includegraphics[width=0.45\textwidth]{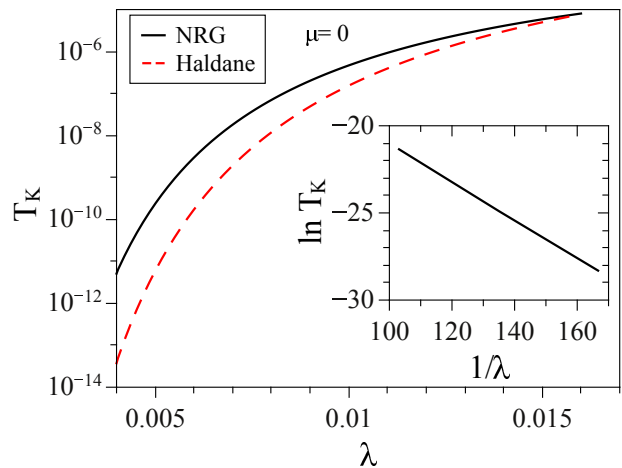}
\caption{\label{fig:TK_mu=0} (Color online)
Kondo temperature $T_K$ vs Rashba parameter $\lambda$ for chemical potential
$\mu = 0$ and hybridization prefactor $\Gamma_0=0.08$, comparing NRG
calculations (solid line) with Haldane's formula Eq.\ \eqref{Haldane} (dashed
line). The inset makes clear the exponential dependence of $T_K$ on $1/\lambda$,
characteristic of a Kosterlitz-Thouless quantum phase transition.}
\end{figure}

We determine the Kondo temperature using the standard operational definition
$T_K \chiimp(T_K) = 0.0701$ based on the universal scaling of the impurity
contribution to the static magnetic susceptibility of the Kondo
model.\cite{Wilson1975}

Figure \ref{fig:TK_mu=0} plots the Kondo temperature $T_K$ as a function of
the Rashba parameter for the case $\mu=0$ where $\Gamma(E)$ takes its minimum
value at $E=\mu$. The figure also shows the prediction 
\begin{equation}
\label{Haldane}
T_K = \sqrt{U\Gamma(\mu)/2} \: \exp\left[-\pi U/8\Gamma(\mu)\right]
\end{equation}
obtained from Haldane's formula \cite{Haldane1978} for the Kondo temperature
of an Anderson model with a flat hybridization function
$\Gamma(E) = \Gamma(\mu) \, \Theta(D-|E|)$.
The inset to Fig.\ \ref{fig:TK_mu=0} establishes the exponential dependence of
$T_K$ on $1/\lambda$. The exponential vanishing $T_K$ as $\lambda\rightarrow 0$
is a signature of a quantum phase transition of Kosterlitz-Thouless type
\cite{Hofstetter2001, Vojta2002, Wong2012} at $\lambda=0$. That the NRG yields
a larger $T_K$ than given by the Haldane formula, especially at low $\lambda$,
is the result of the true hybridization function [Eq.\ \eqref{Gamma}]
satisfying $\Gamma(E) > \Gamma(\mu)$ for all $E\ne 0$.

\begin{figure}[tb]
\includegraphics[width=0.45\textwidth]{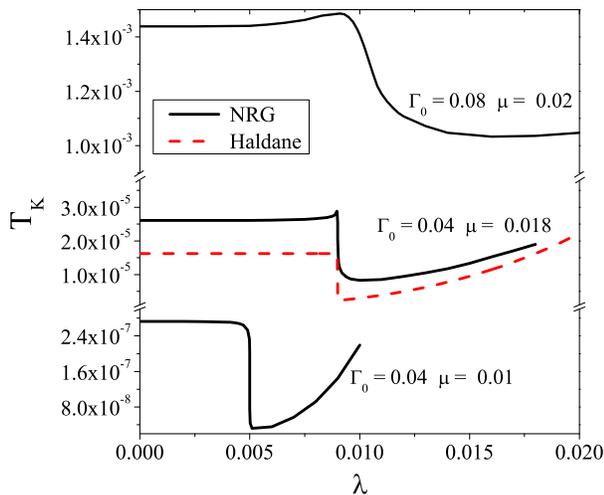}
\caption{\label{fig:Tk_vs_lambda_fixed_mu} (Color online)
Kondo temperature $T_K$ vs Rashba parameter $\lambda$ for three different
combinations of the hybridization prefactor $\Gamma_0$ and the chemical
potential $\mu$ that, for $\lambda=0$, take the system from its mixed valence
regime [largest value of $\Gamma(\mu)$, top case] to deep in its Kondo
regime [smallest $\Gamma(\mu)$, bottom case]. Solid lines plot NRG results,
while the dashed line shows the prediction of Eq.\ \eqref{Haldane} for the
middle value of $\Gamma(\mu)$.}
\end{figure}

Figure \ref{fig:Tk_vs_lambda_fixed_mu} shows examples of the variation of
$T_K$ with $\lambda$ away from the charge-neutrality point. Data are presented
for three different combinations of $\Gamma_0$ and $\mu$. In each case, the
Kondo temperature is almost constant as the Rashba parameter increases from
zero until there is a rapid drop in $T_K$ centered close to $\lambda=|\mu|/2$.
Further increase of $\lambda$ causes $T_K$ to rise and eventually surpass its
value for $\lambda=0$. As shown for the case $\Gamma_0=0.04$, $\mu=0.018$, the
main trends in the NRG results (solid lines) are captured quite well (dashed
line) by Eq.\ \eqref{Haldane} based on
$\Gamma(\mu)=2\Gamma_0|\mu|/D$ for $\lambda<|\mu|/2$
and $\Gamma(\mu)=\Gamma_0(|\mu|+\lambda)/D$ for $\lambda>|\mu|/2$.

Figure \ref{fig:Tk_vs_lambda_fixed_mu} does show some deviations
from the approximation in Eq.\ \eqref{Haldane}. First, just as in the case
$\mu=0$ considered in Fig.\ \ref{fig:TK_mu=0}, the formula systematically
underestimates the Kondo temperature due to its neglect of regions of larger
$\Gamma(E)$ far from the chemical potential. Second, the qualitative shape of
the $T_K(\lambda)$ curve evolves with the degree of electronic correlation,
which can be measured by the ratio $U/\pi\Gamma(\mu) = 2.0$ (top case in Fig.\
\ref{fig:Tk_vs_lambda_fixed_mu}), 4.4, and 8.0 (bottom case). In the most
strongly correlated case, the NRG data show a downward rounding of
$T_K(\lambda)$ for $\lambda$ just below $|\mu|/2$, whereas the other two cases
exhibit a noticeable rise in $T_K$ as $\lambda$ approaches $|\mu|/2$ from
below. These features, as well as a shift of the minimum in $T_K$ to a location
$\lambda>|\mu|/2$, must arise from a subtle balance between increases and
decreases in $\Gamma(E)$ over different decades of $|E-\mu|$.
There is also a progressive smearing of sharp features in $T_K$ vs
$\lambda$ as $U/\pi\Gamma(\mu)$ decreases, signaling a shift from
pure-Kondo behavior (for which use of Haldane's formula is justified)
toward mixed valence (where the formula is inapplicable).

\begin{figure}[tb]
\includegraphics[width=0.43\textwidth]{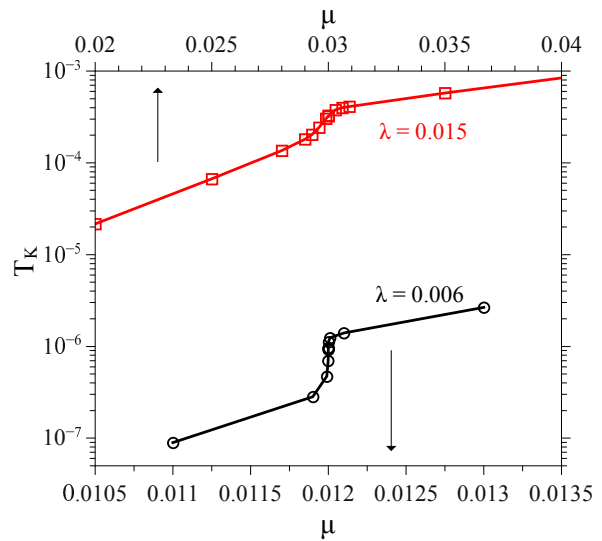}
\caption{\label{fig:TK_vs_mu} (Color online)
Kondo temperature $T_K$ vs chemical potential $\mu$ near the jump in
$\Gamma(\mu)$ at $\mu=2\lambda$, calculated for $\Gamma_0=0.04$ and two
values of the Rashba parameter: $\lambda = 0.006$ (circles, lower axis) and
$\lambda= 0.015$ (squares, upper axis). There is a rapid rise in $T_K$ as
$\mu$ crosses the discontinuity in the hybridization function, with a sharper
variation for the smaller $\lambda$.}
\end{figure}
 
Now we consider variation of the chemical potential at fixed Rashba
parameter. There is a sharp jump $\Delta\Gamma(\mu)=\Gamma(0)\propto\lambda$
as $\mu$ rises through $2\lambda$.
This jump manifests itself in a rapid increase in $T_K$ vs $\mu$, as shown
in Fig.\ \ref{fig:TK_vs_mu}. For $\lambda=0.006$, the Kondo temperature rises
by an order of magnitude as $\lambda$ increases by about 1\%.
For $\lambda=0.015$, the absolute values of $T_K$ and the size of the jump in
$\Gamma(\mu)$ are larger than for $\lambda=0.006$, but the relative increase
in $T_K$ on passing through $\mu=2\lambda$ is only half an order of magnitude.
This can be understood from the fact that $\ln T_K \sim -1/\Gamma$, 
so $\partial\ln T_K/\partial\ln\Gamma\sim 1/\Gamma\sim 1/\lambda$, meaning
that the change in $T_K$ due to the sharp variation of $\Gamma$ around
$\mu=2\lambda$ is softened for increasing $\lambda$.

\subsection{Thermodynamic and spectral quantities}

We now turn to the variation with temperature $T$ of static impurity
thermodynamic properties and to the frequency variation of $T=0$ dynamical
quantities, considering situations where the chemical potential is close
to a point where $\Gamma(\mu)$ jumps in value from $3\Gamma(0)$ to
$4\Gamma(0)$. We focus on $\mu=2\lambda$, although the results would be
identical for $\mu=-2\lambda$ due to the $p$-$h$ symmetry in the host and
(for $U=-2\epsilon_d$) in the impurity.

\begin{figure}[tb]
\includegraphics[width=0.45\textwidth]{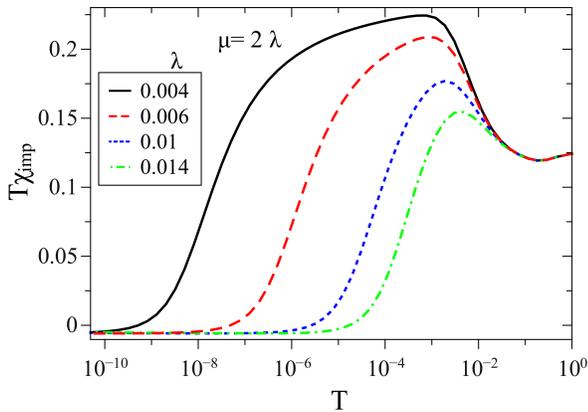}
\caption{\label{fig:Tchi_mu=2lam} (Color online)
Impurity contribution to the magnetic susceptibility at chemical potential
$\mu=2\lambda$, calculated for $\Gamma_0=0.04$ and different values of the
Rashba parameter $\lambda$. At low temperatures, $\chiimp(T)$ changes sign,
showing a residual negative susceptibility contribution as $T\rightarrow 0$.}
\end{figure}

Fig.\ \ref{fig:Tchi_mu=2lam} shows the magnetic susceptibility at
$\mu=2\lambda$, plotted as $T\chiimp$ vs $T$ for different values of the
Rashba parameter ranging from $0.004$ to $0.014$. The curves exhibit features
typical of Kondo screening, with $T\chiimp$ increasing from near $1/8$ 
(the high-temperature susceptibility of the impurity level
when decoupled from the host graphene) for $T\gg U,\,|\epsilon_d|$ toward its 
local-moment value $1/4$ in an intermediate temperature range before falling 
toward zero over four decades of temperature around $T_K$. However, it should 
be noted that $\lim_{T\to 0}T\chiimp(T)$ is not zero, but rather negative. This 
distinctive property is associated with the sharp jump in $\Gamma(E)$, similar 
to behavior found in other systems where the hybridization function has a 
strong energy dependence \cite{Hofstetter1999, Zhuravlev2009, Zhuravlev2011, 
Mastrogiuseppe}.
As $\chiimp$ is the impurity contribution to the susceptibility, the negative
values mean that the introduction of the magnetic adatom lowers the 
susceptibility compared to that of an impurity-free graphene layer.

\begin{figure}[tb]
\includegraphics[width=0.47\textwidth]{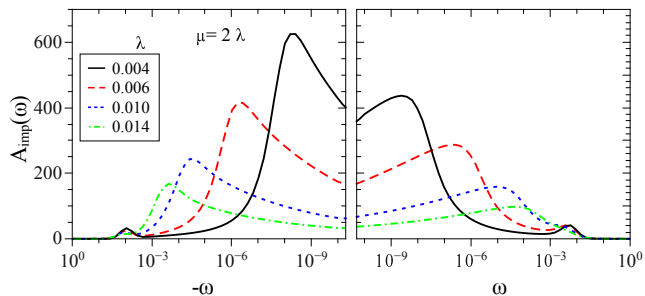}
\caption{\label{fig:Aimp_mu=2lam} (Color online)
Impurity spectral function $A_{\imp}(\omega)$ vs frequency $\omega$ at $T=0$
for $\mu=2\lambda$, $\Gamma_0=0.04$, and different values of the Rashba
parameter $\lambda$. The Kondo peak spanning $|\omega|\lesssim T_K$ is
strongly asymmetric and features a sharp drop around the Fermi level due to
the presence of the jump in $\Gamma(E)$ at $E=\mu$.}
\end{figure}

The impurity spectral function for $\mu=2\lambda$ plotted in Fig.\
\ref{fig:Aimp_mu=2lam} exhibits an asymmetric peak straddling the Fermi
level ($\omega=0$), compatible with a Kondo resonance spanning the window
$|\omega|\lesssim T_K$, but with a sharp dip superimposed that splits the
peak into two parts. As we will see below, a similar feature appears even in
the noninteracting case $U=0$, where it can be traced to the discontinuity in
$\Gamma(E)$ at the chemical potential. As $\lambda$ increases, the combined
``peak-dip'' feature becomes wider, tracking the increase in $T_K$.

\begin{figure}[tb]
\includegraphics[width=0.48\textwidth]{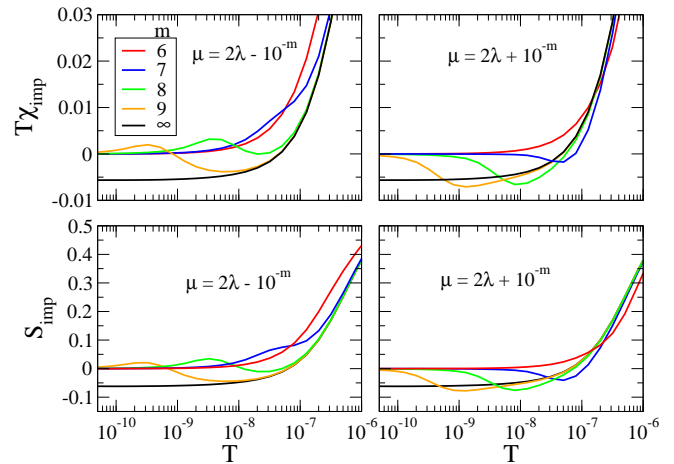}
\caption{\label{fig:thermo_mu=2lam_pm} (Color online)
Low-temperature variation of temperature times the impurity magnetic
susceptibility $T\chiimp$ (upper panels) and impurity entropy $\Simp$
(lower panels) for $\Gamma_0=0.04$, $\lambda =0.006$, and different values of
the chemical potential $\mu = 2\lambda \pm 10^{-m}$ close to a jump in
$\Gamma(\mu)$.}
\end{figure}

\begin{figure}[tb]
\includegraphics[width=0.45\textwidth]{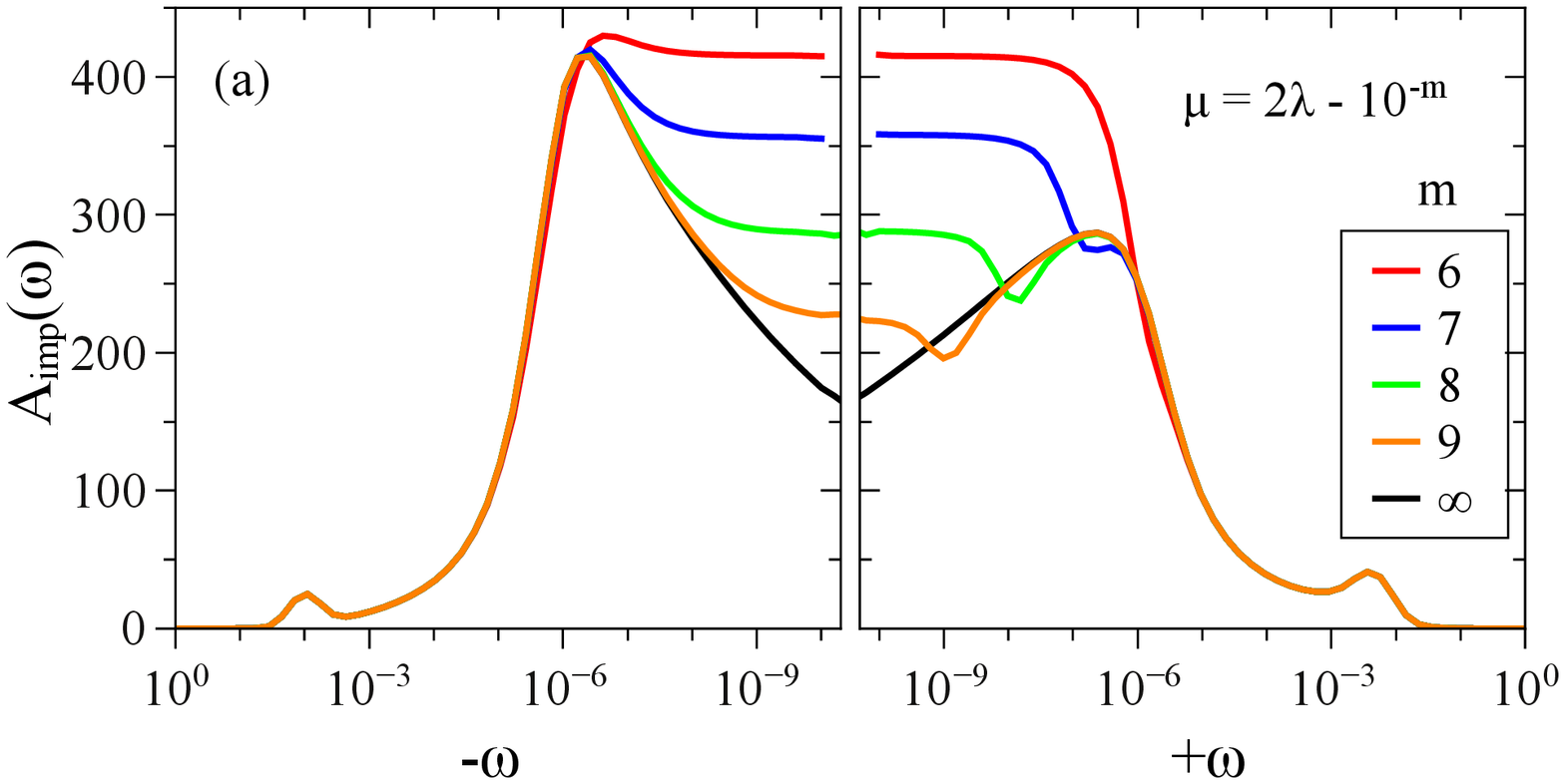}
\includegraphics[width=0.45\textwidth]{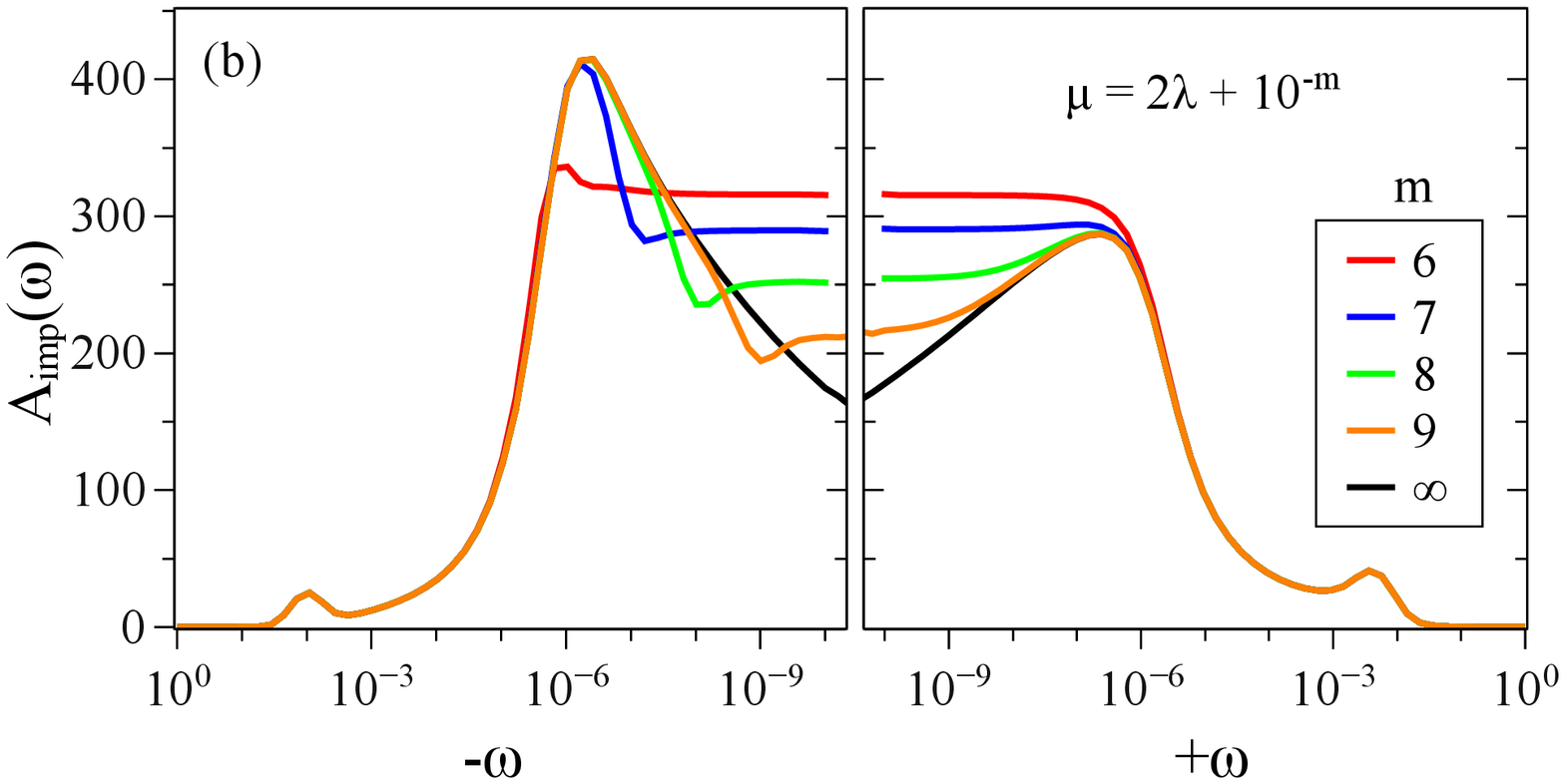}
\caption{\label{fig:Aimp_mu=2lam_pm} (Color online)
Impurity spectral function $A_{\imp}(\omega)$ vs frequency $\omega$ at $T=0$,
calculated for $\Gamma_0=0.04$, $\lambda=0.006$, and different values of the
chemical potential $\mu = 2\lambda \pm 10^{-m}$ close to a jump in
$\Gamma(\mu)$. For $T^*\simeq |\mu-2\lambda| \lesssim T_K$, the Kondo peak
shows a dip near the location $\omega= 2\lambda-\mu$ of the discontinuity.}
\end{figure}

Figures \ref{fig:thermo_mu=2lam_pm} and \ref{fig:Aimp_mu=2lam_pm} show
properties at fixed $\lambda=0.006$ for a set of chemical potentials
$\mu=2\lambda \pm 10^{-m}$ with different integer values of $m$.
When the chemical potential is sufficiently far from the discontinuity
($m\lesssim 6$), both $T\chiimp$ and $\Simp$ show temperature dependences
characteristic of Kondo screening and approach zero monotonically over the
temperature range shown. As $\mu$ gets closer to the jump in $\Gamma(\mu)$,
the low-temperature behavior changes in a way that depends on the sign of
$\mu-2\lambda$. For $\mu<2\lambda$, $T\chiimp$ and $\Simp$ both cross to
negative values and then back to positive values before approaching zero
from above as $T\rightarrow 0$. For $\mu>2\lambda$, by contrast, $T\chiimp$
and $\Simp$ each change sign once and approach zero from below. The properties
deviate from their counterparts for $\mu=2\lambda$ (i.e., $m\to\infty$) below
a characteristic temperature scale $T^*\simeq |\mu-2\lambda|$. The range of
$\mu$ in which unconventional behavior is found is essentially the one in
which $T^*\lesssim T_K$.

The impurity spectral function for $\lambda=0.006$ and $\mu=2\lambda\pm10^{-m}$
is shown in Fig.\ \ref{fig:Aimp_mu=2lam_pm}. For values of $\mu$ far enough
from the discontinuity (as in the case $m=6$), there is a conventional, if
slightly asymmetric, Kondo peak at the Fermi level. Once $\mu$ approaches the
jump close enough that $T\chiimp$ and $\Simp$ undergo sign changes
(i.e., $T^*\simeq|\mu-2\lambda|\lesssim T_K$), the shape of $\Aimp(\omega)$ is
modified around the Fermi level. The peak loses spectral weight, particularly
on the side satisfying $\omega\,(2\lambda-\mu)>0$. The value of $\Aimp(0)$
drops, while $\Aimp(T_K)$ and $\Aimp(-T_K)$ change more slowly, resulting in
splitting of the Kondo peak into two asymmetric parts around a minimum at
$\omega\simeq(2\lambda-\mu)\simeq T^* \, \text{sgn}(2\lambda-\mu)$. 

In order to gain insight into the nature of the unusual behavior reported
above for $\mu$ near a discontinuity in the hybridization function, it is
useful to consider the case $U=0$, for which the impurity spectral function
may be expressed as \cite{Hewson}
\begin{equation}
 \Aimp(\omega) = \frac{1}{\pi} \frac{\Gamma(\omega)}
    {[\omega - \epsilon_d - \Sigma'_d(\omega)]^2 + \Gamma(\omega)^2},
\end{equation}
with $\Gamma(\omega)$ given by Eq.\ \eqref{Gamma} and
\begin{equation}
\begin{split}
 \Sigma'_d(\omega) &= \frac{1}{\pi} P \int \! d\omega'
   \frac{\Gamma(\omega')}{\omega-\omega'}\\
 &= \frac{V^2 \Omega_0}{4 \pi v_F^2} \Biggl\{(\omega+\mu)
    \biggl[2\log\bigg|\frac{(\omega+\mu)^2-4\lambda^2}{(\omega+\mu)^2-1}\bigg|\\
 &- \log{\bigg|1-\frac{4\lambda^2}{(\omega+\mu)^2}\bigg|} \bigg]
  - \log{\bigg|\frac{\omega+\mu-2\lambda}{\omega+\mu+2\lambda}\bigg|} \Biggr\}.
\end{split}
\end{equation}

\begin{figure}[tb]
\includegraphics[width=0.45\textwidth]{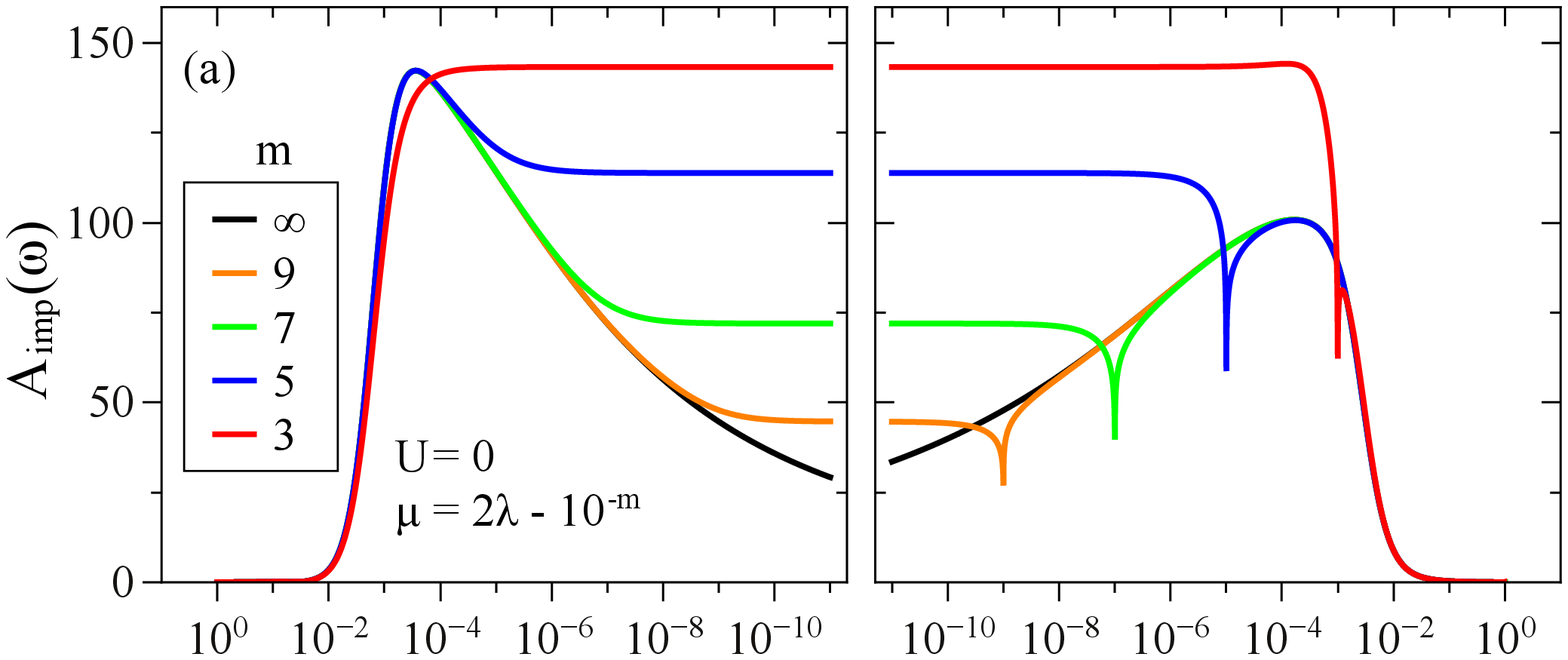}
\includegraphics[width=0.45\textwidth]{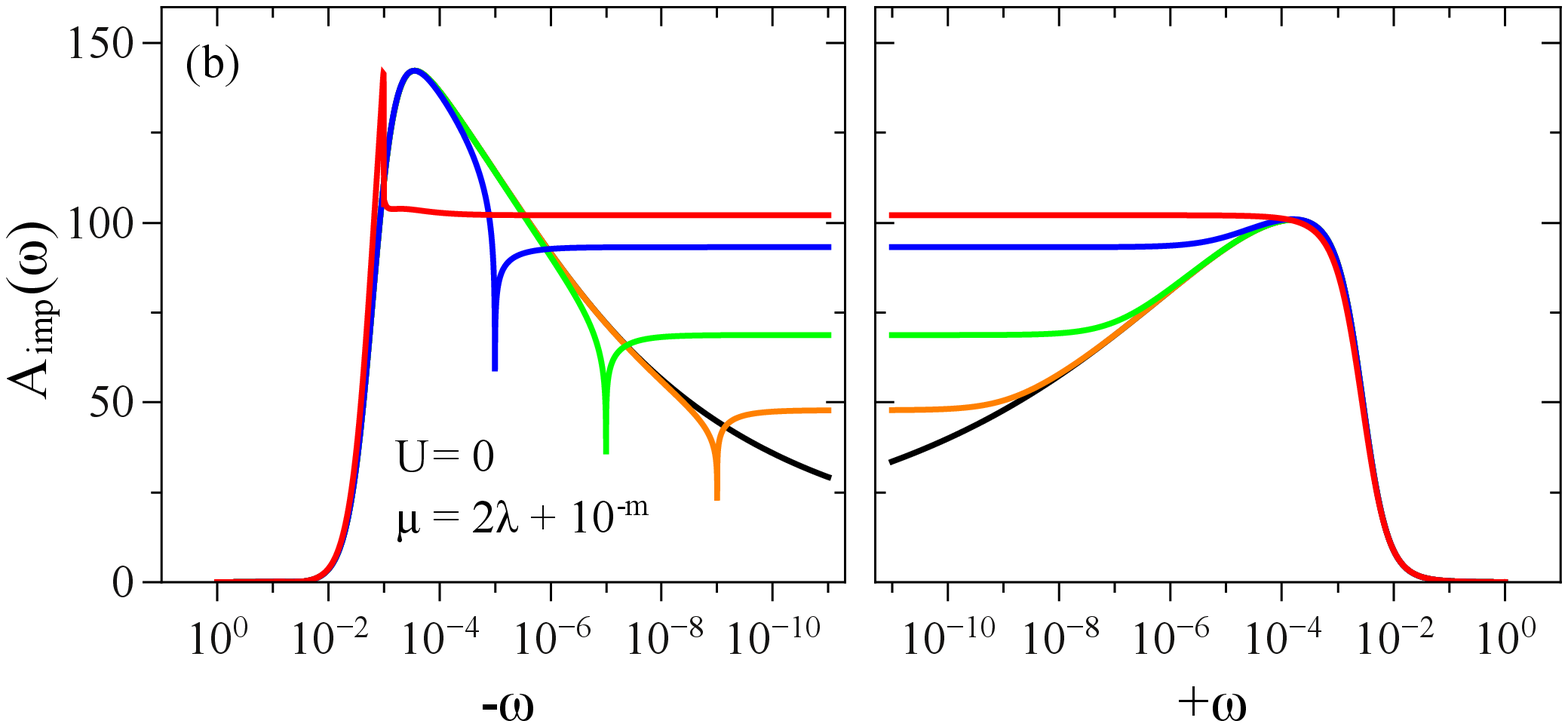}
\caption{\label{fig:Aimp_U=0} (Color online)
Impurity spectral function $\Aimp$ vs frequency $\omega$ at $T=0$ for the
noninteracting case $U=0$ with $\epsilon_d=0.007$, $\Gamma_0=0.04$,
$\lambda=0.015$, and different values of the chemical potential
$\mu = 2\lambda \pm 10^{-m}$ close to a jump in $\Gamma(\mu)$. A sharp,
asymmetric dip structure is centered near the location $\omega= 2\lambda-\mu$
of the discontinuity in the hybridization function.}
\end{figure}

Figure \ref{fig:Aimp_U=0} shows the noninteracting spectral function
$\Aimp(\omega)$ for $\lambda = 0.015$ and $\mu=2\lambda\pm 10^{-m}$.
The impurity level energy is set to $\epsilon_d=0.007$ in order to have a
resonance in the spectral function located near the Fermi level, playing the role
analogous to a Kondo peak in the interacting case. Comparing this figure with the
interacting one in Fig.\ \ref{fig:Aimp_mu=2lam_pm}, one observes many
qualitative similarities. However, the dip in $\Aimp(\omega)$ near the position
$\omega=2\lambda-\mu$ of the hybridization function discontinuity appears much
sharper for $U=0$ than in the interacting case, due to rescaling introduced by 
the interactions and NRG discretization effects.

One feature that is quantitatively similar between Figs.\
\ref{fig:Aimp_mu=2lam_pm} and \ref{fig:Aimp_U=0} is the asymmetry in the
value of $\Aimp(0)$ under reversal in the sign of $\mu-2\lambda$. For $\mu$
fairly far from the discontinuity in $\Gamma(\mu)$ (e.g., $m=6$ in the figures),
the value of $\Aimp(0)$ is one-third higher for $\mu<2\lambda$ than for
$\mu>2\lambda$. Both for $U=0$ and for $U>0$, this can be understood in terms of
approach to the flat-band limit in which the Friedel sum rule gives
$\Aimp(0)=1/[\pi\Gamma(\mu)]$. For smaller values of $|\mu-2\lambda|$, the
nontrivial variation of $\Gamma(E)$ near $E=\mu$ modifies the form of the
Friedel sum rule \cite{DSI06} and leads to $\Aimp(0)<1/[\pi\Gamma(\mu)]$.

We conclude from this comparison with the case $U=0$ that the principal
features of the impurity spectral functions shown in Fig.\
\ref{fig:Aimp_mu=2lam_pm} stem from the jump in the hybridization
function. The dip or antiresonance in $\Aimp$ at the location of the
hybridization step is reminiscent of behavior found previously in systems
with a gapped DOS \cite{Chen1998,Galpin2008a,Galpin2008a,Moca2010,
Mkhitaryan2013,Mastrogiuseppe}, where resonances of single-particle
character can appear inside the gap due to the jump onset in the
hybridization function at the the gap edge.

\section{Discussion}
\label{sec:discuss}

We have presented a study of the Kondo screening of a nondegenerate magnetic
impurity adatom on graphene in the presence of Rashba spin-orbit interaction.
The impurity has been assumed to sit on top of a carbon atom, shown by recent
experiments to be the most likely position for Co impurities. This
configuration can be described by an Anderson impurity model in which a
localized level mixes with a single effective band via an energy-dependent
hybridization function $\Gamma(E)$ that is directly proportional to the graphene
density of states and contains a pair of jumps whose magnitude is proportional
to the Rashba strength $\lambda$, located at energies $E=\pm 2\lambda$.

We have analyzed different regimes that can be accessed by tuning the chemical
potential $\mu$ and the Rashba strength $\lambda$. At $\mu=\lambda=0$ (only),
$\Gamma(\mu)=0$ and hence for $\Gamma_0<\Gamma_{0,c}$ the system lies in a
free-moment phase where the ground state contains a decoupled impurity spin.
$\Gamma_{0,c}=\infty$ for the case $U=-2\epsilon_d$ considered in our
calculations, but away from strict $p$-$h$ symmetry ($U\ne-2\epsilon_d$, not
shown) the critical hybridization prefactor $\Gamma_{0,c}$ would be finite
(Ref.\ \onlinecite{GBuxton1998}).
For $\Gamma_0>\Gamma_{0,c}$ or for any nonzero value of $\mu$ and/or $\lambda$,
the system instead lies in a strong-coupling (Kondo) phase in which the
impurity degree of freedom is quenched at temperatures much below $T_K$. This
Kondo scale is exponential in $-1/\Gamma(\mu)$, indicating that the singular
point $\mu=\lambda=0$ is the location of a Kosterlitz-Thouless type of quantum
phase transition.

When the chemical potential lies close to one of the jumps in $\Gamma(E)$, the
impurity contributions to the static magnetic susceptibility and the entropy
show unusual behavior with decreasing temperature, including a sign change (or
even two) before Kondo screening ultimately sets in. Similar features have been
predicted before in other situations where the hybridization function exhibits
rapid or discontinuous energy dependence \cite{Hofstetter1999, DSI06,
Zhuravlev2009, Zhuravlev2011,Mastrogiuseppe}. In the same range of $\mu$,
the impurity spectral function shows anomalies connected to those seen in the
thermodynamic quantities. The Kondo peak is asymmetric about $\omega=0$ and
for $|\mu-2\lambda| \lesssim T_K$ has a sharp dip-like structure, which can be
traced back to a similar feature found in the noninteracting limit of the model.

Our results have been derived based on an effective description
$H_0^{\eff}=H_g^{\eff}+H_R^{\eff}$ [Eqs.\ \eqref{H_g^eff} and \eqref{H_R^eff}]
of the host graphene, obtained via a low-order expansion of $H_0=H_g+H_R$
[Eqs.\ \eqref{H_g} and \eqref{H_R}] in powers of $q_x a$ and $q_y a$, where
$\bo{q}=\bo{k}-\bo{K}_{\tau}$ is the deviation in reciprocal space from one
or other of the two Dirac points $\tau=\pm$ found for $\lambda=0$. The
effective description yields the hyperbolic band dispersions given in Eq.\
\eqref{dispersion} and the density of states in Eq.\ \eqref{dos} having a
minimum value $\rho(0)=\Omega_0\lambda/\pi v_F^2>0$. However, it has been
shown \cite{Zarea2009} that a complete analysis of $H_0=H_g+H_R$ yields a
band structure that for $\lambda_R>0$ has six Dirac points, three in each
valley $\tau=\pm$ at reciprocal space locations satisfying
$|\bo{q}|a \simeq 2(\lambda_R/t)^2$. As a result, $\rho(E)$ deviates from
the form given in Eq.\ \eqref{dos} for $|E|<E_1\propto \lambda_R^4/t^3$,
dropping linearly to zero at $E=0$ rather than approaching a nonzero limit.

As a consequence of the behavior $\rho(E)\propto|E|$ for $|E|\lesssim E_1$,
when the system is tuned to half filling ($\mu=0$), the pseudogap condition
$\Gamma(\mu)=0$ holds for all values of $\lambda=3\lambda_R/2$ (not just
for $\lambda=0$ as found using $H_0^{\eff}$). This means that for a
$p$-$h$-symmetric impurity ($U=-2\epsilon_d$), the system always has a
free-moment ground state, while in other cases the free-moment ground state
holds for sufficiently weak impurity-host hybridization.
It is important to note, though, that Kondo physics of an Anderson impurity
in a host described by $H_0$ will differ from that reported in Sec.\
\ref{sec:results} for a system described by $H_0^{\eff}$ only on temperature
scales $T\lesssim E_1$. The impurity moment will in many cases appear to be
quenched for $E_1\lesssim T\lesssim T_K$, where $T_K$ is the effective Kondo
scale deduced using the low-order description $H_0^{\eff}$, and only for
$T\ll E_1$ will the many-body screening unwind to reveal an asymptotically
free local impurity moment. For all physically plausible values of
$\lambda_R$ (smaller than $10$\,meV, say), $E_1$ will be orders of magnitude
below the base temperature $T_{\text{min}}$ of any experiment, and there will
be no detectable difference between results for $H_0$ and those for
$H_0^{\eff}$.

The physics we have described should be accessible through scanning tunneling
microscopy on a graphene sample decorated with a few magnetic adatoms.  To
approach the quantum phase transition between the Kondo and free-moment
phases, one could vary $\lambda$ at half filling by manipulating the substrate
and/or the hydrogenation level, or by applying an electric field while keeping
the graphene charge-neutral. Even though the free-moment phase is confined to
$\mu=\lambda=0$ (at least within the effective description of the host provided
by $H_0^{\eff}$), in a real experiment the system would appear to display
free-moment behavior once $\lambda$ becomes small enough that
$T_K<T_{\text{min}}$. Another interesting regime of the system should be
accessible via gate tuning of the chemical potential close to a jump in the
host density of states. In systems with strong Rashba coupling (such as graphene
on a Ni substrate with Au intercalation \cite{Marchenko2012} or hydrogenated
samples \cite{Balakrishnan2013}), scanning tunneling spectroscopy should allow
one to observe the characteristic dip structure in the spectral function.
Although intercalation or decoration could induce disorder and 
modify the DOS, it has been observed that Au intercalation tends to decouple 
graphene from the Ni substrate, leading to a quasi-pristine graphene structure 
\cite{Marchenko2012}.

In systems with more moderate Rashba coupling, it may be possible to observe a
rapid change in $T_K$ as the chemical potential is varied by a few percent
around $2\lambda$.

In conclusion, we find that Kondo screening in graphene is robust against the
presence of Rashba spin-orbit interaction, even though this coupling breaks the
spin symmetry of the Hamiltonian. Electrons with different projections 0, $\pm 1$
of the total angular momentum about an axis perpendicular to the graphene layer
recombine to form a single effective band of screening fermions. The density
of states of this band has a strong energy dependence that leads to nontrivial
phenomena. Our results suggest experimental signatures that may also
characterize the Kondo physics in the new generation of layered two-dimensional
compounds where spin-orbit interactions plays an even stronger role.

\section*{Acknowledgments}

We thank M.\ Zarea for useful discussions.
This work was supported in part under NSF Materials World Network Grants
No.\ DMR-1107814 (Florida) and No.\ DMR-1108285 (Ohio), as well as by NSF-PIRE
grant  No.\ 0730257. D.M., N.S., and S.E.U.\ acknowledge the hospitality of the
Dahlem Center and support from the A.\ von Humboldt Foundation.

\bibliography{refs_rashba}
\bibliographystyle{apsrev4-1}
\end{document}